# Visualizing Poiseuille flow of hydrodynamic electrons

J.A. Sulpizio[1†], L. Ella[1†], A. Rozen[1†], J. Birkbeck[2,3], D.J. Perello[2,3], D. Dutta[1], M. Ben-Shalom[2,3,4], T. Taniguchi[5], K. Watanabe[5], T. Holder[1], R. Queiroz[1], A. Stern[1], T. Scaffidi[6], A.K. Geim[2,3], and S. Ilani[1*]

[1] *Department of Condensed Matter Physics, Weizmann Institute of Science, Rehovot 76100, Israel.*
[2] *School of Physics & Astronomy, University of Manchester, Manchester M13 9PL, United Kingdom.*
[3] *National Graphene Institute, University of Manchester, Manchester M13 9PL, United Kingdom.*
[4] *Department of Physics & Astronomy, Tel-Aviv University, Israel.*
[5] *National Institute for Materials Science, 1-1 Namiki, Tsukuba, 305-0044 Japan.*
[6] *Department of Physics, University of California, Berkeley, California, 94720, USA*
[†] These authors contributed equally to the work.
[*] Correspondence to: shahal.ilani@weizmann.ac.il

Hydrodynamics is a general description for the flow of a fluid, and is expected to hold even for fundamental particles such as electrons when inter-particle interactions dominate. While various aspects of electron hydrodynamics were revealed in recent experiments, the fundamental spatial structure of hydrodynamic electrons, the Poiseuille flow profile, has remained elusive. In this work we provide the first real-space imaging of Poiseuille flow of an electronic fluid, as well as visualization of its evolution from ballistic flow. Utilizing a scanning nanotube single electron transistor, we image the Hall voltage of electronic flow through channels of high-mobility graphene. We find that the profile of the Hall field across the channel is a key physical quantity for distinguishing ballistic from hydrodynamic flow. We image the transition from flat, ballistic field profiles at low temperature into parabolic field profiles at elevated temperatures, which is the hallmark of Poiseuille flow. The curvature of the imaged profiles is qualitatively reproduced by Boltzmann calculations, which allow us to create a 'phase diagram' that characterizes the electron flow regimes. Our results provide long-sought, direct confirmation of Poiseuille flow in the solid state, and enable a new approach for exploring the rich physics of interacting electrons in real space.

The notion of viscosity arises in hydrodynamics to describe the diffusion of momentum in a fluid under the application of shear stress. When scattering between constituent fluid particles becomes dominant, viscosity manifests as an effective frictional force between fluid layers. The hallmark of such hydrodynamic transport in a channel is a parabolic, or Poiseuille, velocity flow profile, which typifies familiar phenomena like water flowing through a pipe. Electron flow has long been predicted[1] to undergo hydrodynamic transport when the rate of momentum-conserving Coulomb scattering between electrons exceeds that of momentum-relaxing scattering from impurities, boundaries and phonons[2–4]. The implications of a dominant viscous force on electronic flow have been studied in wide range of theoretical works[5–10]. While initial efforts were primarily based on linearized Navier-Stokes equations, which describe electron hydrodynamics in the context of diffusive transport[11–13], there is now a developing understanding that a central part of the physical picture is the emergence of hydrodynamics from ballistic flow[14–22]. Reaching the hydrodynamic regime in experiment requires materials of such high purity that the influence of ohmic, transport can be minimized, which is now possible in a growing number of high-mobility systems. Indeed, recent experiments have demonstrated the existence of negative non-local resistance[20,21], superballistic flow[14], signatures of Hall viscosity[23,24], breakdown of the Wiedemann-Franz law[25,26], and anomalous scaling of resistance with channel width[27], which are all phenomena associated with hydrodynamic electron flow. Yet, the real-space observation of the fundamental Poiseuille flow profile has remained elusive.

In this work, we provide the first real space imaging of Poiseuille flow of hydrodynamic electrons, as well as the evolution from ballistic to hydrodynamic flow. We utilize our recently developed technique that employs a scanning carbon nanotube single electron transistor (SET) to non-invasively image real-space maps of the longitudinal and Hall voltage of electrons flowing through high-mobility graphene/hBN channels[28]. By varying the carrier density and temperature, we tune the two relevant length scales that control the electron flow: the momentum relaxing mean free path, set by electron-impurity and electron-phonon scattering, and the momentum conserving mean free path, set by electron-electron interactions. We find that the spatial profile of the Hall field across the channel is a key physical quantity to distinguish the evolution from ballistic into hydrodynamic flow. At low temperatures, we observe flat profiles associated with ballistic flow. At higher temperatures the profiles become parabolic, with curvature approaching that of ideal Poiseuille flow. Overall, we find that Boltzmann kinetic equations qualitatively reproduce our observations, although at the highest temperatures they consistently underestimate the curvature of the Hall field profiles. Finally, we show that this curvature is the distinctive metric for characterizing the different flow regimes, allowing us to construct a phase diagram and map the regions explored by the experiment.

The studied devices are high-mobility graphene/hBN heterostructures patterned into channels of various lengths, $L$, and widths, $W$. Below we present data from a device with $W = 4.7 \mu m$ and $L = 15 \mu m$ (Fig. 1a), but similar results have been obtained for devices with different widths, aspect ratios, and etched boundaries (Supp. Info. S4).

We first perform the scanning analogue of transport measurements of longitudinal resistivity, $\rho_{xx}$. Flowing current $I$ through the channel and imaging the potential produced by the flowing electrons, $\phi(x)$, along the centerline of the channel (dashed line fig. 1b), yields $\rho_{xx} = W \frac{d\phi}{dx} / I$. Fig 1c shows $\rho_{xx}$, measured in this fashion, as a function of the perpendicular magnetic field, $B$, for various carrier densities, $n$, at temperature $T = 7.5$K. Notably, with increasing $n$, $\rho_{xx}$ evolves from a single- to double-peaked structure. This is a well-known signature of ballistic bulk transport, appearing when the momentum-relaxing mean free path, $l_{MR}$, is larger than $W$ and scattering at the walls is diffuse[26,29,30]. As expected, the $B$ dependence of $\rho_{xx}$ is set by the ratio of $W$ and the cyclotron radius, $R_c = \frac{\hbar \sqrt{\pi n}}{eB}$ ($\hbar$ is the reduced Planck's constant and $e$ is the electron charge), as shown by plotting the measurements as a function of $W/R_c$ (top x-axis). For $|W/R_c| > 2$, the cyclotron orbits become smaller than $W$, strongly suppressing backscattering. At these fields, Boltzmann theory predicts[15] that $\rho_{xx}$ is determined primarily by bulk scattering (up to a correction $\sim \left| \frac{W}{R_c} \right|^{-1}$), allowing us to estimate the bulk mean free path, $l_{MR}$ (details in Supp. Info S1). Fig. 1d plots the extracted $l_{MR}$ as a function of $n$ at several different temperatures.

For $T = 7.5$K, $l_{MR}$ exhibits the expected $\sqrt{n}$-dependence, while at $T = 75$K and $150$K, $l_{MR}$ displays a characteristic flat density dependence due to the addition of phonon scattering, consistent with previous estimates[30].

Next, we image the potential of the flowing electrons[28] along the transverse ($y$) direction (dashed line, fig. 2a) perpendicular to the current flow, at $B = 0$, to evaluate the spatial resolution of the SET imaging. In the absence of a Lorentz force, the potential should be constant as a function of $y$, dropping sharply to zero at the etched walls. The imaged potential, plotted as a function of the normalized coordinate, $\frac{y}{W}$ (Fig. 2b blue), is indeed flat in the bulk of the channel, but has a rounded drop at the walls. This rounding reflects our spatial resolution, set by the height separation between the SET and the graphene ($h \approx 880nm$ in the current experiment) and is accurately reproduced (dashed yellow) by convolving a flat-top potential with our point spread function[28]. All subsequent analysis is thus based only on the bulk of the channel ($|y/W| < 0.3$, grayed regions near walls are ignored), where the effect of smearing at the walls into the channel is negligible.

We now turn to the Hall voltage profiles, which are fundamentally related to the current flow profiles of electrons in the channel. In the ohmic regime ($l_{MR} \ll W$), there is a local relation between the y-component of the Hall field, $E_y = \frac{dV_{Hall}}{dy}$, and the current density parallel to the channel axis, $j_x$, given by $E_y = \frac{B}{ne}j_x$. In the hydrodynamic regime, in which the electron-electron mean free path, $l_{ee}$, is smaller than the size of the sample, the current density is predicted to be parabolic, leading to an analogous relation[31] with the Hall field (Supp. Info. S6):

$$E_y = \frac{B}{ne}\left(j_x + \tfrac{1}{2}l_{ee}^2\partial_y^2 j_x\right). \tag{1}$$

Deep in the hydrodynamic regime, where $l_{ee}/W \ll 1$, the second term becomes small and the local relation between $E_y$ and $j_x$ is recovered to a good approximation. Imaging $E_y(y)$ in these regimes therefore effectively images the current distribution, $j_x(y)$. In the ballistic regime, this local relation breaks down, leading to a fundamentally different $E_y$ profile. As we show below, this makes $E_y$ a key observable for distinguishing between ballistic and hydrodynamic flow. Figure 2c shows the potential along $y$ measured at small magnetic fields $B = \pm 12.5\text{mT}$, anti-symmetrized in $B$, to yield the Hall voltage profile $V_{\text{Hall}}(y) = \phi(y, B) - \phi(y, -B)$, ($T = 7.5K$, $n = -1.5 \times 10^{11}\text{cm}^{-2}$). Note that $B$ is small enough so that the flow remains semiclassical (Landau level filling factor $\nu \gg 100$ and $\hbar \omega_c \ll k_b T$ $\omega_c$ is the cyclotron frequency). Below we obtain the $E_y(y)$ profiles by numerically differentiating such $V_{\text{Hall}}(y)$ profiles.

We now observe how electron-electron interactions affect the flow profiles by comparing the Hall field imaged at different temperatures: $T = 7.5K$ in fig. 2e, and $T = 75K$ in fig. 2f. Note that while increased temperature is expected to increase the electron-electron scattering rate (decrease $l_{ee}$) it is also known to increase the electron-phonon scattering (decrease electron-phonon mean-free-path, $l_{ph}$) and correspondingly reduce $l_{MR} = \left(l_{imp}^{-1} + l_{ph}^{-1}\right)^{-1}$, where $l_{imp}$ is the impurity scattering mean free path. To make the best comparison that isolates the influence of electron-electron interactions, we therefore maintain a nearly constant $l_{MR}$ across the different temperatures by tuning the carrier density between the measurements (circles, Fig. 2d). Notably, the imaged profile at $T = 7.5K$ is flat across the bulk of the channel (Fig. 2e). In contrast, the profile at $T = 75K$ is strongly parabolic (Fig. 2f). The dramatic difference in curvature between these profiles becomes even more apparent when we image the full 2D maps of the Hall field (within the black square in fig 2a), demonstrating that the shape of the profiles does not depend on a specific position along the channel (fig 2g,h). We note that although a nonzero magnetic field is needed to produce a measurable Hall voltage in these measurements ($W/R_c = 1.3$), we demonstrate experimentally in Supp. Info. S2 that this field is small enough as to minimally influence the profile. Additionally, we show in Supp. Info. S3 that the voltage excitation $V_{ex}$ applied across the channel is sufficiently low to not induce electron heating.

One naively expects the current density profile, $j_x(y)$, to be flat for ballistic flow and parabolic for hydrodynamic, Poiseuille flow. However, a full Boltzmann theoretical calculation of the profiles of $j_x$ and $E_y$ which includes the effect of $l_{MR}$ (Fig. 2i,j and supp.

info S5) leads to two surprising conclusions that deviate from this expectation: First, the $j_x$ profile, even deep in the ballistic regime ($l_{MR}/W \gg 1$), is not flat. Fig. 2i plots the $j_x$ profile calculated for $l_{MR}/W = 2$ and $l_{ee}/W = 4.3$, consistent with our measurements at $T = 7.5K$, showing that $j_x$ still has significant curvature. In fact, the Boltzmann theory predicts a significantly curved $j_x$ profile even for much larger $l_{MR}/W$, and reveals that such ballistic $j_x$ profiles only become flat logarithmically in $l_{MR}/W$, which would require an unphysically long $l_{MR}$ to observe in experiment. The curvature of the ballistic $j_x$ profile is therefore not qualitatively different from the curvature in the hydrodynamic regime (e.g. calculated for $l_{MR}/W = 1.4$ and $l_{ee}/W = 0.16$ in fig 2j) and is therefore a weak marker for the emergence of electron hydrodynamics from ballistic flow. Secondly, Boltzmann theory shows that the profile of $E_y$, in contrast to that of $j_x$, differs markedly between ballistic and hydrodynamic flows: In the ballistic regime, the $E_y$ profile is indeed flat for the parameters in our measurement (fig 2i) and can even acquire a negative curvature if $l_{MR}/W$ is increased further (see measurements below). In the hydrodynamic, Poiseuille regime, the theory predicts a positively curved $E_y$ profile (fig. 2j). This establishes that the curvature of the $E_y$ profile is a key quantity for distinguishing between ballistic and hydrodynamic electron flow.

The $E_y$ profile in fig. 2j is calculated to best fit our measurements at $T = 75K$ (fig 2f) using a Knudsen number of $K_n \equiv l_{ee}/W = 0.16$. This is consistent with hydrodynamic electron flow in which $l_{ee}$ is indeed the smallest length scale in the system, and is in agreement with previous transport measurements[20,23]. Comparing the $j_x$ and $E_y$ profiles calculated for these parameters (fig 2j), we see that they are similarly curved, although not identical. This is consistent with equation (1) above relating $E_y$ to $j_x$ in the hydrodynamic regime, where the proportionality between these quantities is restored with a correction $\sim (l_{ee}/W)^{-2}$. In the measurement in fig. 2f, the observed $E_y$ profile thus approximates the actual Poiseuille $j_x$ profile to within 5% (right y-axis). Note that the theoretical $j_x$ profile corresponding to the $T = 75K$ measurement does not reach zero at the walls, allowing us to estimate a slip length[32] of $l_{slip} \sim 500nm$

Having imaged the emergence of Poiseuille flow at increased temperatures, we now explore the dependence of the electron-electron interactions on carrier density. Following a basic prediction of Fermi liquid theory for a linearly dispersing spectrum, $l_{ee} \sim \frac{E_F}{T^2} \sim \frac{\sqrt{n}}{T^2}$ ($E_F$ is the Fermi energy), a variation of the flow profiles with $n$ is also expected. Varying $n$, however, will generically also change $l_{MR}$, possibly masking the relatively weak effects due to the $\sqrt{n}$-dependence of $l_{ee}$. Fortunately, at elevated temperatures there is a wide range of $n$ over which $l_{MR}$ remains nearly constant due to compensating effects of phonon and impurity scattering (between the green dots in fig 3a, at $T = 75K$). In fact, examining the magnetoresistance at two different densities (green dots in fig 3a), we see that the curves are nearly identical for all values of $B$ (green curves in fig 3b), implying that from transport measurements alone it would be impossible to distinguish between electron flows at these densities. However, the corresponding imaged $E_y$ profiles (green curves, fig 3c) are markedly different, varying in curvature by ~50%, which reflects the variation in $l_{ee}$. This result highlights again that the $E_y$ profile is a sensitive indicator for hydrodynamics. At even lower $n$ (black dot, fig 3a) $l_{MR}$ drops and we observe changes both in the measured magnetoresistance (fig 3b, black) and the imaged $E_y$ profiles (fig 3c, black) as compared to higher densities.

We now systematically investigate how the curvature of $E_y$ varies over a broader range of $n$ and $T$. For each $n$ and $T$ we image the $E_y$ profile, fit to the form $E_y(y) = ay^2 + c$ for $|y/W| < 0.3$, and extract the normalized curvature defined as $\kappa = -\frac{a}{c}\left(\frac{W}{2}\right)^2$. This definition is chosen such that $\kappa = 0$ corresponds to a flat profile and $\kappa = 1$ corresponds to an ideal parabolic Poiseuille profile that reaches zero at the walls. Figure 4a plots the measured $\kappa$ as a function of $n$ for $T = 7.5K, 75K$ and $150K$. At $T = 7.5K$ we find that $\kappa$ is close to zero, and even becomes negative at high density. We further observe that the value of $\kappa$ monotonically increases with increasing $T$ and decreasing $n$, with the measured curvature approaching the ideal Poiseuille value at the highest $T$ and $n$.

As seen above, the curvature of the $E_y$ profiles nicely captures the different electron flow regimes. To demonstrate this more concretely, we plot in fig. 4b a phase diagram of the flow based on $\kappa$ calculated using the Boltzmann theory as a function of the two length scales that control the physics: $l_{MR}/W$ and $l_{ee}/W$. To quantitatively compare to the experiment, we include in the calculations the finite $W/R_c = 1.3$ and the small correction due to the smearing induced by the PSF of our imaging. Notably, the value of $\kappa$ demarcates four regions in phase space: ohmic, ballistic, Poiseuille, and porous, the last two of which are hydrodynamic. In the ohmic regime (lower right quadrant), $l_{ee}/W \gg 1$ while $l_{MR}/W < 1$, with $\kappa$ reaching zero for the smallest values of $l_{MR}/W$. Increasing $l_{MR}/W$ beyond 1 while maintaining $l_{ee}/W \gg 1$, a transition between the ohmic and ballistic regimes occurs. In contrast with the simplistic view in which the profile is flat whenever $l_{ee}/W \gg 1$, there is a peak in curvature when $l_{MR}/W \approx 0.25$, with the curvature reaching a maximal value $\kappa_{max} = 0.31$. In the ballistic regime, where $l_{MR}/W \gg 1$, the curvature is governed by the reciprocal sum $\left(\frac{1}{l_{ee}} + \frac{1}{l_{MR}}\right)^{-1}$, and even becomes negative. Such negative curvature is consistent with our measurements at $T = 7.5K$ (fig. 4a), and can be shown to result from ballistic effects reminiscent of magnetic focusing. In the left half of the phase diagram ($l_{ee}/W < 1$), the flow is hydrodynamic, which is either Poiseuille (top left) or porous (bottom left) in character. The transition occurs when the so-called 'Gurzhi' length, $D_\nu = \frac{1}{2}\sqrt{l_{ee}l_{MR}}$, crosses through $W$. In the porous regime ($D_\nu < W$), named in analogy to water flow through porous media, both $l_{MR}$ and $l_{ee}$ can be smaller than $W$. In this regime, the curvature in the bulk of the channel is low as in the ohmic regime, but electron-electron interactions lead to a sharp drop of $E_y$ at the walls. In the Poiseuille regime ($D_\nu > W$), $\kappa$ increases significantly, approaching $\kappa = 1$, with the parabolic profiles of both $E_y$ and $j_x$ reaching zero at the walls (full comparison to $j_x$ curvature in Supp. Info S7).

We now quantitatively compare the imaged $E_y$ profiles at each $n$ and $T$ against the Boltzmann theory. For each $n$ and $T$, using the $l_{MR}$ measured in fig. 1d, we find the $l_{ee}$ that gives the best fit of the Boltzmann profile to the imaged one. The extracted values of $l_{ee}$ (solid lines in fig 4a inset) are in close agreement with the many-body calculation for single layer graphene[14] (dashed lines), exhibiting the predicted decrease of $l_{ee}$ with decreasing $n$ and increasing $T$. We note that once $l_{ee}$ exceeds the length of the channel (dashed black line) the Boltzmann calculations, which assume an infinite channel, lose their predictive power. We also note that while at $T = 7.5K$ and 75K the Boltzmann profiles fit very well both the overall magnitude and the curvature of the imaged $E_y$ profiles, at $T = 150K$, the best fit profiles consistently underestimate the curvature imaged experimentally. This is likely due to the overly simplistic approximation of the scattering integral used in the calculation. This suggests that an improved microscopic understanding of electron-electron interactions, beyond the scattering time approximation used here and in most Boltzmann treatments of electrons hydrodynamics, is necessary to more completely understand hydrodynamics in real electronic systems (e.g. using scattering integrals that better account for energy momentum conservation in 2D as proposed in ref[33,34] ). Finally, we overlay the values of $l_{MR}$ and $l_{ee}$ obtained from the measurements onto fig 4b (colored paths correspond to the different temperatures, dots indicate lowest densities), showing the trajectories through the phase diagram explored in the experiment.

In conclusion, we have imaged the flow of electrons through graphene devices using a scanning SET to map the transverse component of the Hall electric field, which we find to be the essential element for distinguishing hydrodynamic from ballistic flow. At the lowest temperatures in the ballistic regime, we image Hall field profiles that are nearly flat. As the temperature is increased, we observe the transition from ballistic into hydrodynamic flow through $E_y$, which develops the characteristic parabolic profile. Because of the convergence of $E_y$ to $j_x$ in the hydrodynamic regime, these images constitute the first experimental observation of Poiseuille electron current profiles, and firmly establish the existence of an electron liquid that flows according to a universal hydrodynamic description. In combination with Boltzmann calculations, we show that the curvature of the profiles defines a phase diagram of the various electron flow regimes. These experiments demonstrate the crucial insights provided by spatial imaging of electron flow, which upon application to other materials and topologically distinct flow geometries, should  enable further exploration of the physics of strongly interacting electrons.

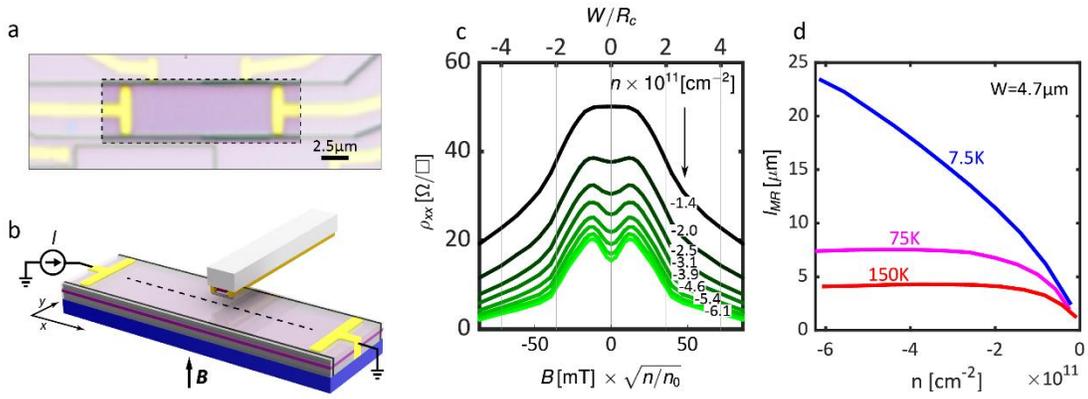

**Figure 1: Overview of graphene channel device and imaging of magnetoresistance**. **a**. Optical image of graphene channel device used for imaging electron flow, consisting a high-mobility, single layer of graphene sandwiched between hBN layers (purple), and electrical contact electrodes (yellow) on top of conducting Si/SiO back gate (blue). The dark lines are etched walls that define a channel of width $W = 4.7\mu m$ and length $L = 15\mu m$ (outlined with the dashed box, scale bar 2.5$\mu m$). **b**. Rendering of scanning SET imaging performed in experiments. The nanotube-based SET is positioned at the end of a scanning probe cantilever, and is rastered across the channel (graphene in purple, sandwiched between hBN layers atop an Si/SiO$_2$ substrate in blue) to locally image the potential generated by the electrical current $I$ in perpendicular magnetic field $B$. **c**. Magnetoresistance of graphene channel at temperature $T = 7.5$K imaged non-invasively with scanning SET. The SET is scanned along the centerline of the channel (black dashed line in panel b) to image the potential drop $\Delta\phi$ in order to extract the longitudinal resistance $\rho_{xx} = \frac{W\frac{\Delta\phi}{\Delta x}}{I}$ as a function of magnetic field (bottom $x$-axis in units of militesla, normalized by $n_0 = -6.1 \times 10^{11}$cm$^{-2}$, top $x$-axis in units of $\frac{W}{R_c} \sim B$, see text) for different charge carrier densities $n$ (numbers labeling each curve, low density in black, high density in green). The high-density magnetoresistance curves show a double-peak structure, indicating ballistic transport. **d**. Momentum-relaxing mean-free path $l_{MR}$ in the bulk of the graphene channel as a function of carrier density for several temperatures. The value of $l_{MR}$ is deduced from $\rho_{xx}(B)$ as described in Supp. Info. S1.

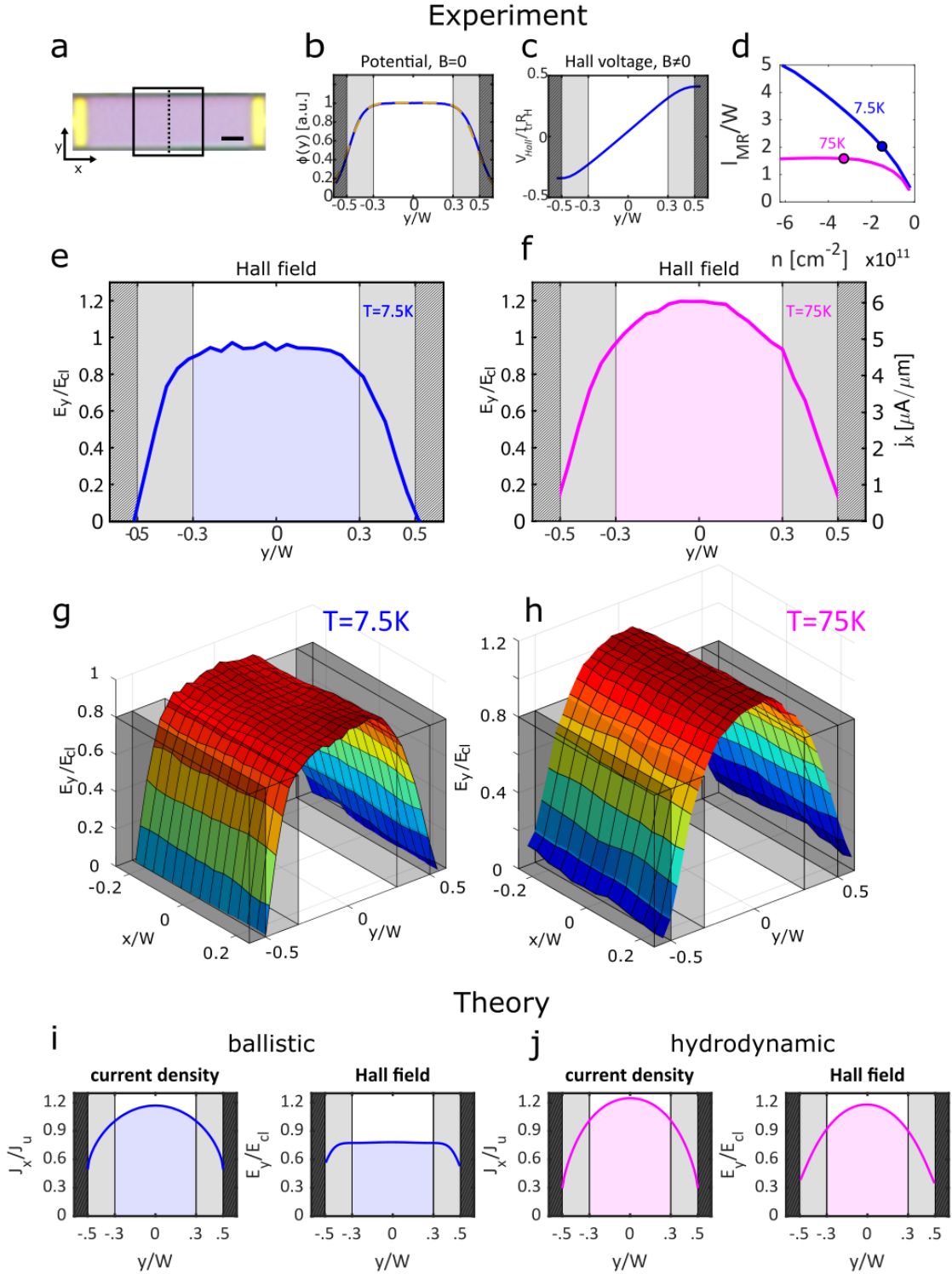

**Figure 2: Imaging ballistic and Poiseuille electron flow profiles**. **a**. Graphene channel with overlay indicating region over which flow profiles are imaged. 1D profiles are taken along the dashed line, 2D profiles are imaged across the region enclosed by the black square (scale bar 2.5μm). **b.** Potential of flowing electrons, $\phi$, as function of the $y$ coordinate (dashed line, panel a) imaged at $B = 0$ (blue, $T = 7.5$K). Dashed yellow

curve is a boxcar function convolved with the point spread function of our measurement, determined primarily by the height of our SET detector above the graphene during the scan. Grayed regions ($0.3 < \left|\frac{y}{W}\right| < 0.5$) indicate where the smearing of the steps at the edges due to the finite spatial resolution has a non-negligible contribution. **c.** Imaged Hall voltage, $V_H$, at field $B = 12.5\text{mT}$, $n = -1.5 \times 10^{11}\text{cm}^{-2}$ and $T = 7.5\text{K}$. **d.** $l_{MR}$ from fig. 1a, but now normalized by $W$. Dots indicate the carrier densities of the profile imaging in all subsequent panels, where $n = -1.5 \times 10^{11}\text{cm}^{-2}$ at 7.5K and $-3.1 \times 10^{11}\text{cm}^{-2}$ at 75K, chosen such that $l_{MR}$ is nearly equal for both temperatures. **e.** The Hall field, $E_y$, at $T = 7.5\text{K}$, normalized by the classical value $E_{\text{cl}} = \left(\frac{B}{ne}\right) I/W$, obtained by numerical differentiation of $V_H$ with respect to $y$. **f.** $E_y$ at $T = 75\text{K}$. The right $y$-axis converts the field to units of current density by scaling with $ne/B$. **g.** 2D map of the $E_y$ taken over the dashed square region in panel a at $T = 7.5\text{K}$. **h.** 2D map of $E_y$ at $T = 75\text{K}$. **i,j.** Calculation of the current density $j_x$ (normalized by $j_u = I/W$) and the Hall field $E_y/E_{\text{cl}}$ based on the Boltzmann theory with values of $l_{MR}$ and $l_{ee}$ corresponding to the experimental data in panels e and f. In panel i, the values used are $\frac{l_{MR}}{W} = 2$ and $\frac{l_{ee}}{W} = 4.3$, whereas for panel j $\frac{l_{MR}}{W} = 1.4$ and $\frac{l_{ee}}{W} = 0.16$. The calculated profiles are convolved with the PSF for direct comparison with the experiment. The current density appears parabolic in both the hydrodynamic and ballistic regimes, whereas the $E_y$ profile is relatively flat in the ballistic regime and parabolic in the hydrodynamic regime.

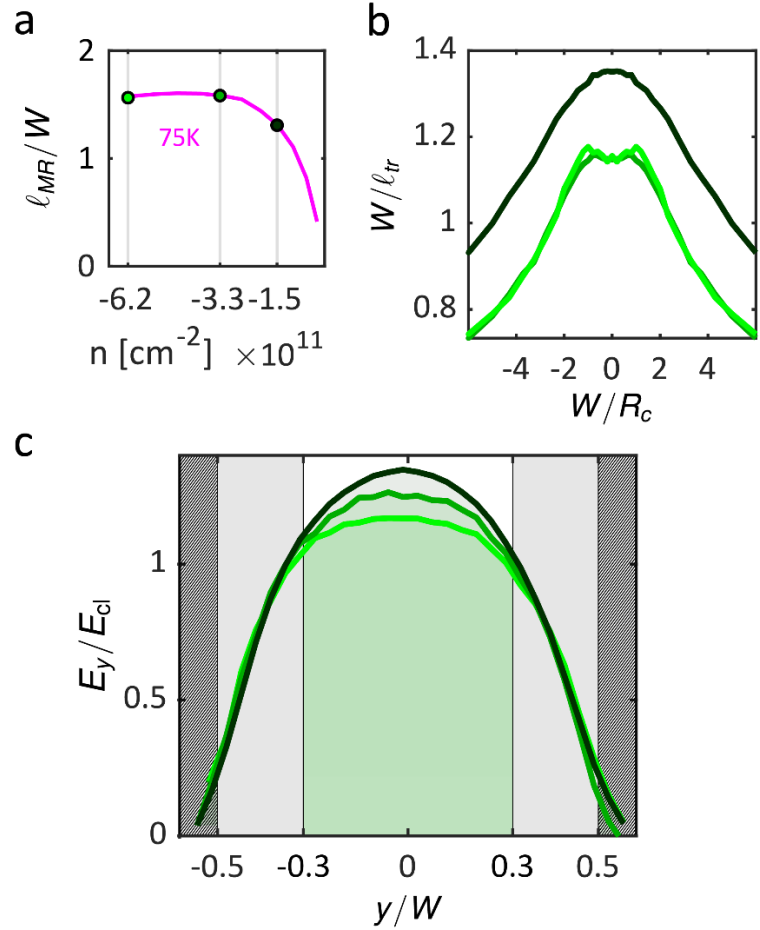

**Figure 3. Carrier density dependence of hydrodynamic electron flow profiles**. **a**. $\frac{l_{MR}}{W}$ for $T = 75K$ taken from fig 1d with dots indicating values of $n$ corresponding to experiments in subsequent panels. Between the green dots, $l_{MR}$ is practically independent of $n$ due to the combination of phonon and impurity scattering. **b**. Comparison of magnetoresistance in units of the inverse mean free path $\frac{W}{l_{tr}}$ at $T = 75K$ for several values of $n$ indicated by the color of the curve (corresponding to dots in panel a). The two green curves at higher $n$ exhibit nearly indistinguishable magnetotransport. **c**. $E_y/E_y^{cl}$ profiles imaged for the same values of $n$ as in panel b as indicated by color, demonstrating the monotonic increase of curvature with decreasing $n$. Here, $E_y^{cl}$ is the bulk value for classical Hall field, $\frac{BI}{neW}$.

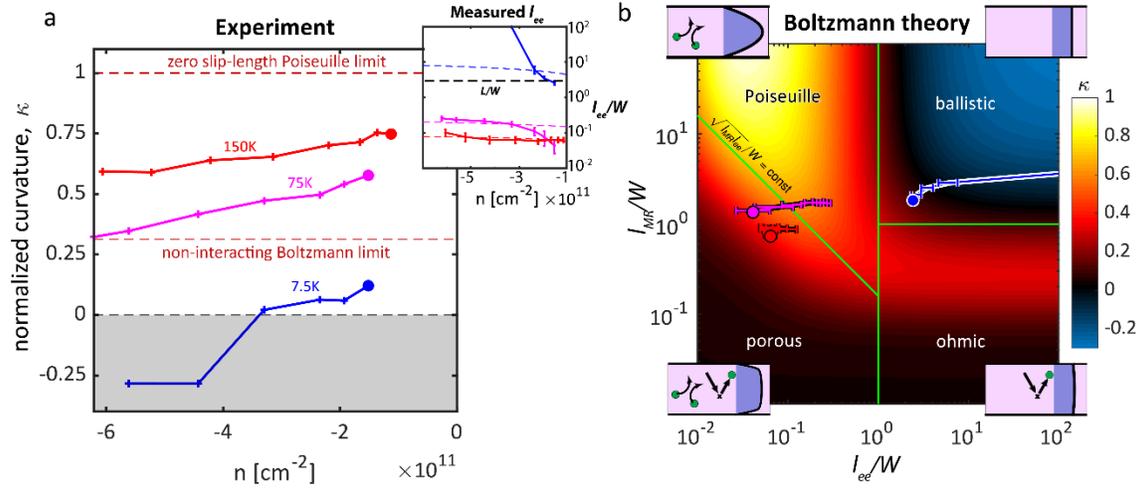

**Figure 4. Curvature of the imaged $E_y$ profiles and phase diagram of electron flow regimes**. **a.** Curvature, $\kappa$, of the imaged $E_y$ profiles as a function of $n$ and $T$ as described in the main text. Dashed red lines mark the maximal curvature obtained for non-interacting electrons based on Boltzmann calculations, and also the curvature of ideal Poiseuille flow with zero slip length. Inset: $l_{ee}$ at the values of $n$ and $T$ from the experiment (solid lines with error bars), determined by comparing the imaged $E_y$ profiles to those calculated using the Boltzmann equations. The colored dashed lines are the corresponding predications for $l_{ee}$ based on many-body calculations for single layer graphene (see Ref [14]). The black dashed line marks the length of the device $L$ (normalized by $W$), above which the Boltzmann theory for an infinitely long channel can no longer predict $l_{ee}$. **b.** Phase diagram of electron flow as obtained from $\kappa$, calculated by Boltzmann theory (colormap) as a function of $\frac{l_{MR}}{W}$ and $\frac{l_{ee}}{W}$. The curvature values are determined after convolving the calculated profiles with the PSF of the experiment and using the same finite magnetic field as in the experiment ($\frac{W}{R_c} = 1.3$). The different electron flow regimes are labeled (ballistic, ohmic, Poiseuille, and porous) together with illustrations of the relevant scattering mechanism. Electrons are drawn as green circles, and $E_y$ profiles are schematically drawn in purple. In the ballistic regime, the $E_y$ profile is flat or even negatively curve. In the ohmic regime, electrons scatter primarily from impurities/phonons (drawn as x's), and the $E_y$ profile can be gently curved. In the Poiseuille regime, electrons primarily scatter from other electrons leading to a strongly parabolic $E_y$ profile. In the porous regime, both impurity/phonon scattering as well as electron-electron scattering play a dominant role, resulting in an $E_y$ profile that is gently curved in the middle of the channel and reaches zero over a distance $\sim D_\nu = \frac{1}{2}\sqrt{l_{MR}l_{ee}}$ from the walls. The green lines mark the transitions between the different regimes: ballistic to ohmic at $\frac{l_{MR}}{W} = 1$, transition to hydrodynamics $\frac{l_{ee}}{W} = 1$, and transition from Poiseuille to porous at $D_\nu/W \sim 1$. In the Poiseuille regime the profiles can reach maximum curvature of $\kappa = 1$. The overlaid blue, purple, and red paths correspond to the values of $l_{MR}$ and $l_{ee}$ (same error bars as in inset) at $T = 7.5\text{K}, 75\text{K},$ and $150\text{K}$, respectively, from the experimental traces in panel a, with the dots indicating the lowest density.

**Methods:**

Device fabrication: Scanning SET devices were fabricated using a nanoscale assembly technique[35]. The graphene/hBN devices were fabricated using electron-beam lithography and standard etching and nanofabrication procedures[20] to define the channels and evaporation of Pt (main text) and Pd/Au (S4) to deposit contact electrodes.

Measurements: The measurements are performed on multiple graphene devices in two separate, home-built, variable temperature, Attocube-based scanning probe microscopes. The microscopes operate in vacuum inside liquid helium dewars with superconducting magnets, and are mechanically stabilized using Newport laminar flow isolators. A local resistive SMD heater is used to heat the samples under study from $T = 7.5K$ to $T = 150K$, and a DT-670-BR bare chip diode thermometer mounted proximal to the samples and on the same printed circuit boards is used for precise temperature control. The voltage imaging technique employed is presented in reference[28]. Voltages and currents (for both the SET and sample under study) are sourced using a home-built DAC array, and measured using a home-built, software-based audio-frequency lock-in amplifier consisting of 1uV accurate DC+AC sources and a Femto DPLCA-200 current amplifier and NI-9239 ADC. The local gate voltage of the SET is dynamically adjusted via custom feedback electronics employing a least squares regression algorithm to prevent disruption of the SET's working point during scanning and ensure reliable measurements.

The voltage excitations applied to the graphene channels were as follows: 4.3mV at $T = 7.5K$, 7.5mV at $T = 75K$, and 15mV at $T = 150K$, all chosen to not cause additional current heating (S3). The magnetic fields applied ranged between $\pm 100$ mT .

**Acknowledgements:** We thank Gregory Falkovich, Andrey Shytov, Leonid Levitov, Denis Bandurin, and Roshan Krishna-Kumar for helpful discussions, and Alessandro Principi for assistance with the many-body theory. We further acknowledge support from the Helmsley Charitable Trust grant, the ISF (grant no. 712539), WIS-UK collaboration grant, the ERC-Cog (See-1D-Qmatter, no. 647413), and the Emergent Phenomena in Quantum Systems initiative of the Gordon and Betty Moore Foundation.

**Data availability:** The data that support the plots and other analysis in this work are available from the corresponding author upon request.

**Contributions:** J.A.S., L.E., A.R., D.D., and S.I. performed the experiments. J.A.S., L.E., A.R., and S.I. analyzed the data. J.B., D.P., and M.B.-S. fabricated the graphene devices. K.W. and T.T. supplied the hBN crystals. T.S, T.H., R.Q., and A.S. performed theoretical calculations. J.A.S, L.E, and S.I. wrote the manuscript, with input from all authors.

Supplementary materials for:

# Visualizing Poiseuille flow of hydrodynamic electrons

Joseph A. Sulpizio[†], Lior Ella[†], Asaf Rozen[†], John Birkbeck, David J. Perello, Debarghya Dutta, Tobias Holder, Raquel Queiroz, Takashi Taniguchi, Kenji Watanabe, Moshe Ben-Shalom, Ady Stern, Thomas Scaffidi, Andre K. Geim, and Shahal Ilani

## Contents

## S1. Determination of the momentum relaxing mean-free-path from magnetoresistance

For a channel geometry of width $W$, as used in the experiments in this paper, the longitudinal resistance, $\rho_{xx}$, reflects both the bulk resistivity of the graphene as well as scattering from the walls. In order to isolate the contribution from the bulk resistivity and determine the momentum relaxing mean-free-path in the bulk, $l_{MR}$, we make use of the measured magnetoresistance. At any magnetic field we can obtain the transport mean free path from the measured $\rho_{xx}$ via $l_{tr}(B) = \frac{h}{2e^2 k_F \rho_{xx(B)}}$. In the semiclassical regime, the primary influence of a perpendicular magnetic field $B$ is to bend the electron trajectories into cyclotron orbits of radius $R_c = \frac{\hbar \sqrt{\pi n}}{eB}$, where $n$ is the charge carrier density, $e$ is the electron charge, and $\hbar$ is the reduced Planck's constant. At small magnetic fields such that the skipping orbit diameter is larger than the channel width, $|W/R_c| < 2$, electrons can be efficiently backscattered in the bulk and by the walls, and thus $l_{tr}(B)$ contains the effects of both bulk and wall scattering. On the other hand, when $|W/R_c| > 2$, the backscattering from the walls is highly suppressed since a cyclotron orbit emerging from one wall cannot reach the other wall without scattering at least once in the bulk. In this regime the transport mean free path is primarily controlled by the bulk scattering length, $l_{tr} \sim l_{MR}$, with a small correction $\sim |W/R_c|^{-1}$ due to the volume participation ratio of skipping cyclotron orbits. In fact, using Boltzmann calculations of the magnetoresistance we can determine the correction factor over the entire phase space of the experiment. Fig S1 shows the ratio, $l_{tr}/l_{MR}$, calculated using Boltzmann theory (section S5 below), as a function of $l_{ee}/W$ and $l_{MR}/W$ for $\frac{W}{R_c} = 3.2$. By estimating the $l_{ee}$ in our experiments using the $E_y$ measurements and the Boltzmann calculations as in the main text (figure 4a inset), and using $l_{tr}$ as a zeroth order estimate for $l_{MR}$, we can determine from fig S1 the correction factor and obtain from our measured $l_{tr}$ the bulk $l_{MR}$. Note that in the regions of the phase diagram traversed by the experiment (curves in fig. 4b in the main text), the correction factor is rather small and the maximal deviation of $l_{MR}$ from $l_{tr}$ is $\sim 30\%$, so even the naïve estimate, $l_{MR} \sim l_{tr}$, would be already rather accurate.

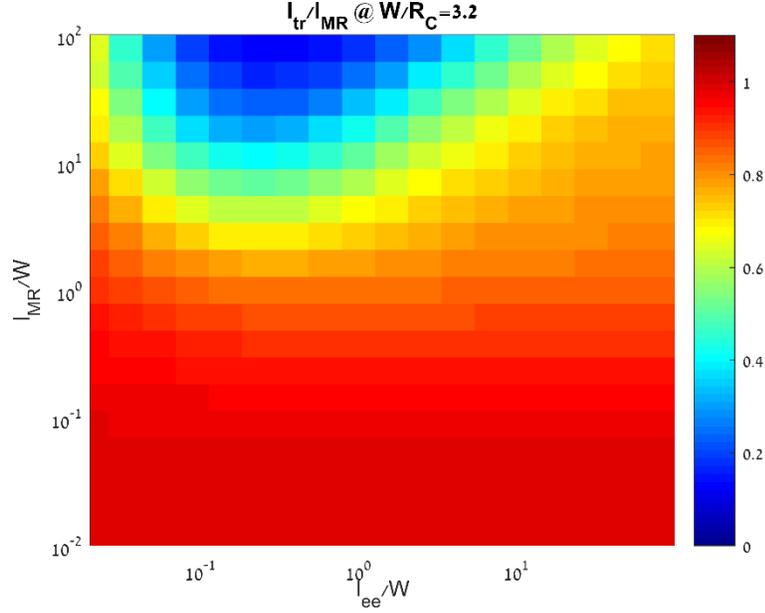

**Figure S1: Relation between transport and bulk mean free path at the skipping orbit regime.** The 2D map shows the ratio of the finite field transport mean free path, $l_{tr}(B) = \frac{h}{2e^2 k_F \rho_{xx(B)}}$, and the bulk mean free path, $l_{tr}/l_{MR}$, calculated using Boltzmann theory at $\frac{W}{R_c} = 3.2$ for a channel with specular walls, as a function of $l_{ee}/W$ and $l_{MR}/W$.

## S2 Dependence of curvature on magnetic field

Our method for mapping the Hall field, $E_y$, relies on the application of a small perpendicular magnetic field, $B$, to produce a Hall signal that is measurable by the scanning SET. We must then verify that this measurement is in the linear response regime with respect to $B$, namely that $B$ is low enough to not alter the $E_y$ profile. Specifically, we aim to prove that the curvature of the $E_y$ profiles, $\kappa$, which is a main observable in this work, is not altered by $B$. In fig S2a, we present $\kappa$ imaged at a constant carrier density as a function of magnetic field at three temperatures, $T = 4\text{K}, 75\text{K}$, and $150\text{K}$. The curvature is extracted as described in the main text by a parabolic fit to $E_y$ over the center of the channel.

We note two distinct regimes of how $\kappa$ depends on $B$: for $\frac{W}{R_c} > 2$, $\kappa$ has a strong field dependence, whereas for $\frac{W}{R_c} < 2$, $\kappa$ is constant at each temperature. In the higher field regime for $\frac{W}{R_c} > 2$, closed cyclotron orbits can fit within the width of the channel. This leads to a rich evolution of $E_y$ profiles which are no longer simply parabolic, which is the topic of a future work. In the lower field regime for $\frac{W}{R_c} < 2$, we see that the measured curvature is constant to within our measurement noise down to lowest fields measured ($W/R_c \sim 1$). Imaging closer to $B = 0$ is increasingly challenging, as the signal-to-noise ratio of the measured Hall voltage decreases linearly with decreasing field. Fig. S2b shows similar traces ($\kappa$ vs $B$) calculated using Boltzmann equations for the values of $\frac{l_{MR}}{W}$ and $\frac{l_{ee}}{W}$ corresponding to the experiment. We find a good correspondence between the Boltzmann simulations and the experiment. Most importantly, in the low field regime for $W/R_c < 2$, the simulations confirm that $\kappa$ is independent of $B$ as observed in the experiments, and extend this observation down to $B = 0$. Based on these results, the value of $\frac{W}{R_c} = 1.3$ used for the $E_y$ profile imaging in the experiments in the main text is justified.

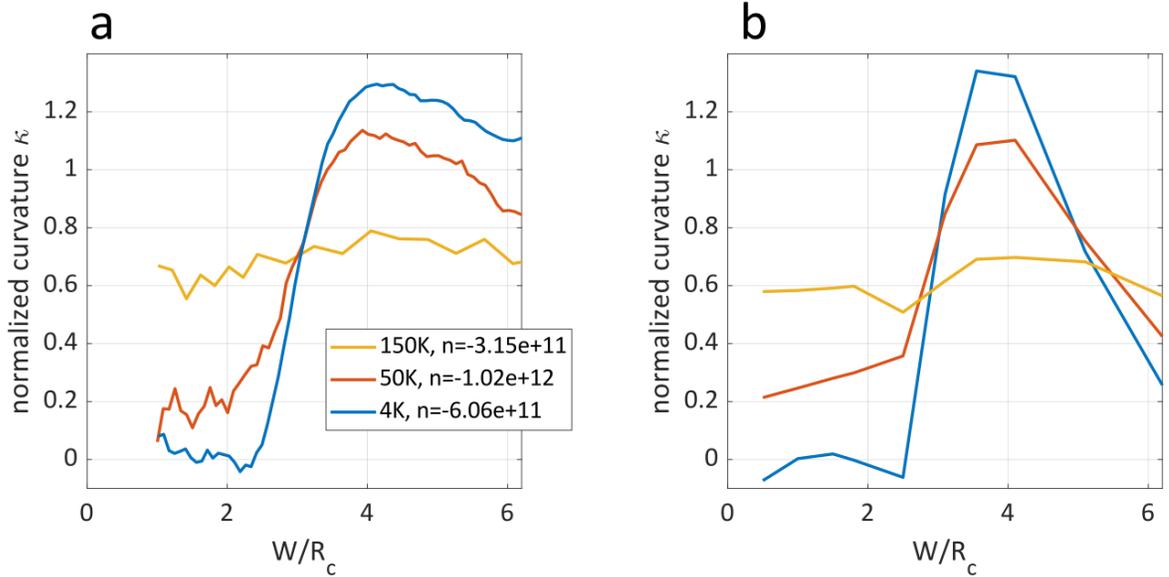

**Figure S2: Dependence of curvature $\kappa$ on applied magnetic field. a.** Measured traces of $\kappa$, extracted from $E_y$ with a fit to the center of the channel, as a function of magnetic field. The field is plotted in units of $\frac{W}{R_c} \propto B$. The blue trace is measured at $T = 4\text{K}$ and hole density of $n = -6.06 \times 10^{11}\text{cm}^{-2}$ on device B (see SI section S4). The orange trace is measured at $T = 50\text{K}$ and $n = -1.02 \times 10^{11}\text{cm}^{-2}$ on device B, and the yellow trace is measured at $T = 150\text{K}$ and a hole density of $n = -3.15 \times 10^{11}\text{cm}^{-2}$ on device A, which is the device used throughout the main text. Two distinct regimes are apparent: Below $W/R_c \simeq 2$ and above. Below this value the curvature is nearly independent of $W/R_c$. **b.** Curvature as a function of $W/R_c$ extracted from a Boltzmann simulation of $E_y$ as described in the main text. Blue trace: $l_{MR}/W = 1.4$, $l_{ee}/W = \infty$. Orange trace: $l_{MR}/W = 1.4$, $l_{ee}/W = 0.14$. Yellow trace: $l_{MR}/W = 1.8$, $l_{ee}/W = 0.18$. As in the experimental traces, an abrupt change in behavior is apparent at $W/R_c \simeq 2$.

# S3. Dependence of curvature on applied voltage excitation

We investigate here the role of the voltage excitation on the curvature of the imaged flow profiles in our experiments. In order to drive current through the graphene channel device, we apply an AC voltage excitation of amplitude $V_{ex}$ between the electrical contacts to the device. This excitation can in principle induce heating of the electrons above the temperature of the cryostat, and as a result cause an increase in curvature of the Hall field profiles. While this effect can be used as in ref [2] instead of substrate heating, we avoid this approach here due to additional spurious effects it may have on the curvature. We therefore choose an excitation amplitude at each temperature that is sufficiently low to minimally influence the curvature of the imaged profiles, but still high enough to enable a robust measurement.

Figure S3 shows the curvature of the field profiles vs excitation amplitude $V_{ex}$ across the graphene device for two temperatures, $T = 7.5\text{K}$ in the ballistic regime (blue) and $T = 75\text{K}$ in the hydrodynamic, Poiseuille regime (purple). The curvature is extracted by a parabolic fit to the imaged $E_y$ hall profile imaged across the channel at a fixed density and magnetic field as described in the main text. In the Poiseuille regime (density $n = -3.3 \times 10^{11}\text{cm}^{-2}$, $\frac{W}{R_c} = 1.3$), we see that the curvature ($\kappa \sim 0.5$) is essentially independent of the excitation at least up to $V_{ex} = 11\text{mV}$, and therefore the excitation does not influence the physics of the electron flow. In the ballistic regime ($n = 1.5 \times 10^{11}\text{cm}^{-2}$, $\frac{W}{R_c} = 1.3$), we see a clear increase in the curvature with increasing excitation due to electron heating.

Still, for an excitation of $V_{ex} = 4.3\text{mV}$, $\kappa$ is nearly zero and far below the Boltzmann limit marking the transition to hydrodynamic flow. We can thus safely choose such a low excitation and robustly image ballistic electron flow through the channel, though the specific value of $\kappa$ may still be somewhat influenced by the excitation. In the experimental data presented in the main text, for $T = 7.5K$, the excitation across the graphene device is chosen such that $V_{ex} < 4.3\text{mV}$, while at higher temperatures, we choose an excitation such that $V_{ex} < 7.5\text{mV}$

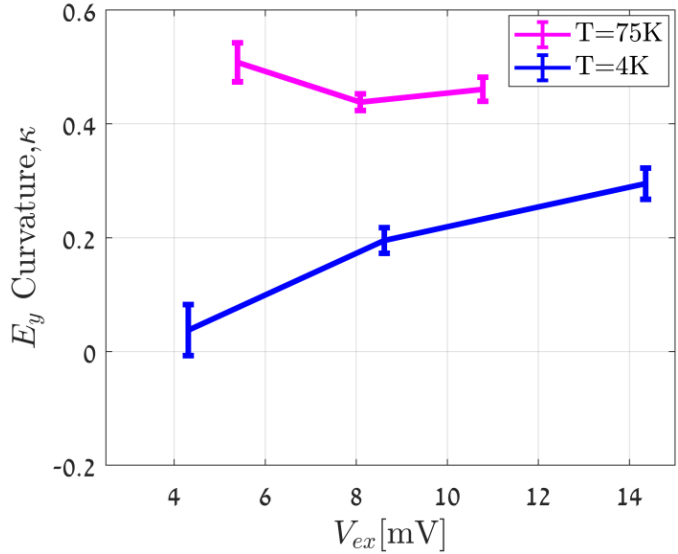

**Figure S3: Dependence of curvature $\kappa$ on bias excitation amplitude**. Normalized curvature $\kappa$ is plotted as a function of the excitation amplitude applied between the contacts of the channel. Blue trace: $T = 4\text{K}$, $n = -1.5 \times 10^{11}\text{cm}^{-2}$. Purple trace: $T = 75\text{K}$, $n = -3.3 \times 10^{11}\text{cm}^{-2}$. This plot verifies that by choosing appropriate values for the excitation as was done for the experiments in the main text, our conclusions are not the result of electron heating effects.

## S4. Measurement of curvature in additional devices

We establish the consistency of our results across a set of graphene channel devices and scanning SET probes. The measurements in this work were carried out on two separate graphene device microchips, each imaged with a different scanning microscope and different SET. This allows us to compare between measurements and establish their lack of sensitivity to details specific to a particular graphene device or experimental setup. We denote the device used throughout the text as device A. The additional device measured, which we denote as device B, is a channel with $W = 5\mu\text{m}$, similar to device A, and $L =$ $42\mu m$, compared to $L = 15\mu m$ in device A. This difference allows us additionally to rule out aspect-ratio dependent effects.

To most easily compare between devices, we examine the curvature of the Hall field profiles imaged at similar SET-graphene device separations. We focus on the magnetic field dependence of the curvature at several different temperatures and densities, similar to section S2 above. The results are shown in fig. S4. We compare first between measurements taken at $T = 7.5$K and $n = -1.5 \times 10^{11} \text{cm}^{-2}$ in device A and $T = 4$K and $n = -6 \times 10^{11} \text{cm}^{-2}$ in device B. We then repeat the same comparison, now at $T = 75$K for both devices and $n = -3.3 \times 10^{11} \text{cm}^{-2}$ for device A and $n = -10 \times 10^{11} \text{cm}^{-2}$ for device B. The point spread function of the SET has a similar influences on both devices, and the same valid channel region is chosen for the extraction of the curvature ($|W/R_c| < 0.3$).

In the low temperature measurement, we observe a similar overall shape in the $W/R_c < 2$ region. The low field curvature in device A levels at slightly higher value than that in device B. The latter can be attributed to the different densities , since, as observed in fig. 4a of the main text, at $T = 7.5$K the curvature exhibits strong density dependence. The curvature imaged at elevated temperature imaging closely match each other over the full range of magnetic fields, with small residual differences that are consistent with the density dependence in fig. 4a of the main text. This indicates that the hydrodynamic features observed in this work are not specific to the particular graphene or channel dimensions being measured.

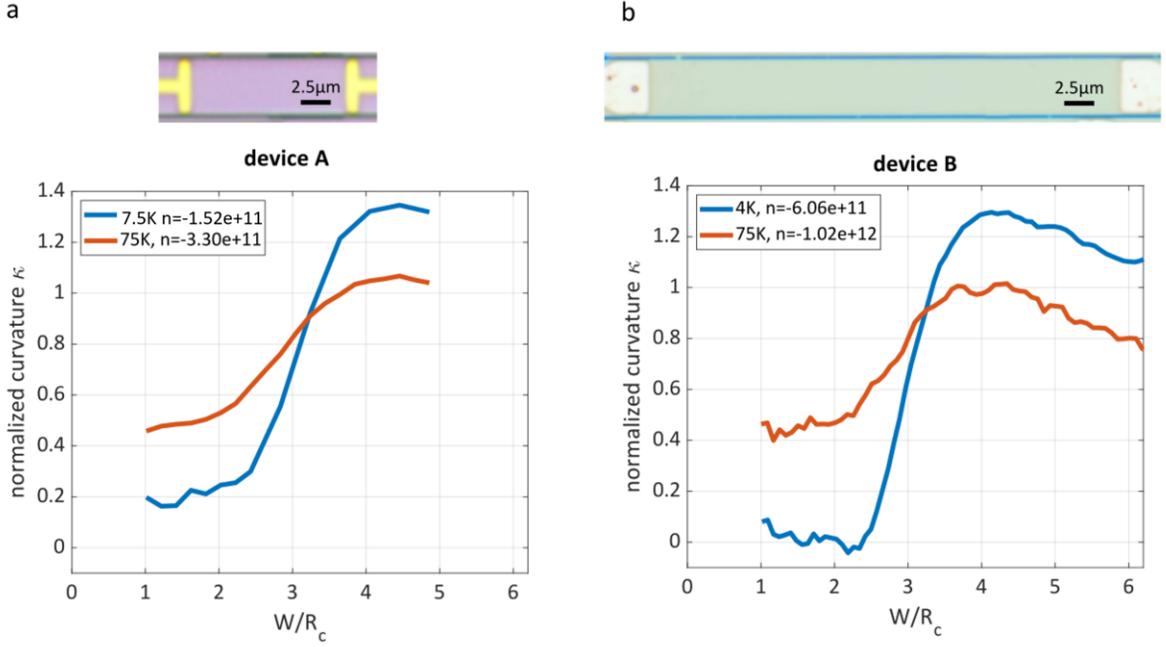

**Figure S4: Comparison to other devices**. **a**. Top: optical image of graphene device (device A) patterned into the geometry of a channel, with $W = 5\mu m$ and $L = 15\mu m$, studied in the main text. Bottom: normalized curvature $\kappa$ as a function of $W/R_c$, compare to section S2. Legend: Temperature and carrier density of measurements. **b**. Top: optical image of an additional graphene device (device B) used for similar measurements, with $W = 5\mu m$ and $L = 42\mu m$. This device was measured in a separate cryostat with a different scanning microscope and different SET. Color differences between optical images are due to variable lighting conditions. Bottom: $\kappa$ vs. $W/R_c$ for device B, showing a result highly consistent with that of panel a.

## S5. Boltzmann simulations of flow profiles

To model electron flow through the graphene channels, we employ an approach based on the Boltzmann equation[2–4] that incorporates the effects of both electron-impurity and electron-phonon scattering as well as electron-electron interactions:

$$\partial_t f + \vec{v} \cdot \nabla_{\vec{r}} f + \frac{e}{m}\left(\vec{E} + \vec{v} \times \vec{B}\right) \cdot \nabla_{\vec{v}} f = \frac{\partial f}{\partial t}\bigg|_{scatt}, \qquad \textbf{(Eq. S3)}$$

where the scattering integral,

$$\frac{\partial f(\vec{r},\vec{v})}{\partial t}\bigg|_{scatt} = -\frac{f(\vec{r},\vec{v}) - n(\vec{r})}{\tau} + \frac{2}{\tau_{ee}}\vec{v} \cdot \vec{j}(\vec{r}), \qquad \textbf{(Eq. S4)}$$

has two contributions: one from momentum-relaxing scattering, with a rate $\frac{1}{\tau_{MR}}$, and one from momentum-conserving, electron-electron scattering, with a rate $\frac{1}{\tau_{ee}}$. This equation describes the evolution of the semiclassical occupation number $f(\vec{r}, \vec{v})$ for a wave packet at position $\vec{r}$ and velocity $\vec{v}$, where $n(\vec{r}) = \langle f \rangle_{\vec{v}}$ is the local charge density, $\vec{j}(\vec{r}) = \langle f \vec{v} \rangle_{\vec{v}}$ the local current density, $\langle ... \rangle_{\vec{v}}$ is the momentum average, and $\frac{1}{\tau} = \frac{1}{\tau_{MR}} + \frac{1}{\tau_{ee}}$. For the sake of simplicity, we consider the case of a circular Fermi surface with $\vec{v} = v_F \hat{\rho}$, where $\hat{\rho}$ is the radial unit vector. Mean free paths are then simply defined as $l_{MR(ee)} = v_F \cdot \tau_{MR(ee)}$. The term proportional to $\tau_{ee}^{-1}$ is the simplest momentum-conserving scattering term that can be written, assuming that the electrons relax to a Fermi-Dirac distribution shifted by the drift velocity[5,6].

We assume a sample that is of infinite length along the $x$-axis (which is the direction of current flow), and of finite width $W$ along the $y$-axis. The magnetic field is applied along the $z$ direction. Diffuse scattering at the boundaries is imposed by the following boundary condition:

$$f\left(y = \pm\frac{W}{2}, \vec{v}_{\mp}\right) = \pm f_{boundary}, \qquad \textbf{(Eq. S5)}$$

where $\vec{v}_{+/-}$ corresponds to all velocities with a positive/negative component along the $y$-axis, and where $f_{boundary}$ is determined self-consistently.

Equation (S1) is supplemented by Gauss's law with a charge density given by $en(\vec{x})$. The resulting integrodifferential equation is solved numerically using the method of characteristics[6] to invert the differential part of the equation, and an iterative method to solve the integral part.

## S6. Relation between $E_y$ and $j_x$ in the hydrodynamic regime

In the hydrodynamic regime for a channel of width $W$ and bulk resistivity $\rho_{xx}$ with diffusive walls, the Hall field $E_y(y)$ across the channel at weak magnetic field $B$ calculated using the Boltzmann kinetic equation approach[7] is given by:

$$E_y(y) = \rho_H j_x - \frac{\frac{E_x 2 l_{ee}}{R_c} \cosh\left(\frac{2y}{D_v}\right)}{\cosh\left(\frac{W}{D_v}\right)}, \qquad \textbf{(Eq. S6)}$$

where $\rho_H = \frac{B}{ne}$ is the Hall resistivity, $j_x$ is the current density along the channel, $E_x$ is the electric field along the channel, $R_c$ is the cyclotron radius, $l_{ee}$ is the electron-electron scattering length, and $D_v = \sqrt{l_{ee} l_{MR}}$ is the Gurzhi length, where $l_{MR}$ is the momentum relaxing mean-free-path. Additionally, we calculate the corresponding current density as:

$$j_x(y) = \frac{E_x}{\rho_{xx}} \left( 1 - \frac{\cosh\left(\frac{2y}{D_v}\right)}{\cosh\left(\frac{W}{D_v}\right)} \right), \qquad \textbf{(Eq. S7)}$$

where $\rho_{xx}$ is the longitudinal resistivity. We then note the following identity:

$$\partial_y^2 j_x = -\frac{E_x}{\rho_{xx}} \left(\frac{2}{D_v}\right)^2 \left( \frac{\cosh\left(\frac{2y}{D_v}\right)}{\cosh\left(\frac{W}{D_v}\right)} \right). \qquad \textbf{(Eq. S8)}$$

This allows us to substitute Eq. (S8) into (S6), using the relation $\rho_H = \frac{\rho_{xx} l_{MR}}{R_c}$, we find:

$$E_y = \rho_H \left( j_x + \frac{1}{2} l_{ee}^2 \partial_y^2 j_x \right). \qquad \textbf{(Eq. S9)}$$

## S7. Theoretical comparison of $E_y$ and $j_x$ throughout the phase diagram

We compare the phase diagram defined by the theoretical estimate for the curvature $\kappa$ of the $E_y$ profiles presented in the main text figure 4b with the phase diagram defined by the theoretical curvature of the $j_x$ current density profiles. This allows us to present a more complete relation between $E_y$ and $j_x$ for $\frac{W}{R_c} = 3.2$ for each flow regime as a function of $\frac{l_{MR}}{W}$ and $\frac{l_{ee}}{W}$. The phase diagrams are presented side-by-side in figure S7. Similar to the $E_y$ curvature phase diagram, the $j_x$ curvature phase diagram is constructed by fitting a parabola to the center of the $j_x$ profiles calculated from the Boltzmann model after convolution with the PSF of the SET. Examining first the right, non-hydrodynamic half of the phase diagram, we note the significant difference between the curvature in the ballistic regime of $E_y$ and $j_x$. Whereas $E_y$ can be negatively curved, $j_x$ is always positively curved, with significant curvature throughout the ballistic regime. At low magnetic fields, this curvature is due to the finite $\frac{l_{MR}}{W}$, while at low values of $\frac{l_{MR}}{W}$, it is due to the non-zero magnetic field. The crossover between the ballistic regime and the ohmic regime is evident in both phase diagrams, although the $j_x$ curvature simply decreases from ballistic to ohmic, while $E_y$ goes through a local maximum near the crossover. In the hydrodynamic regime, as in the ohmic regime, both phase diagrams are highly similar, with the curvature matching exactly in both limits of strongly Poiseuille and porous hydrodynamic electron flow. This highlights the restoration of a local relation between $E_y$ and $j_x$, which leads to a convergence between these quantities in the hydrodynamic regime.

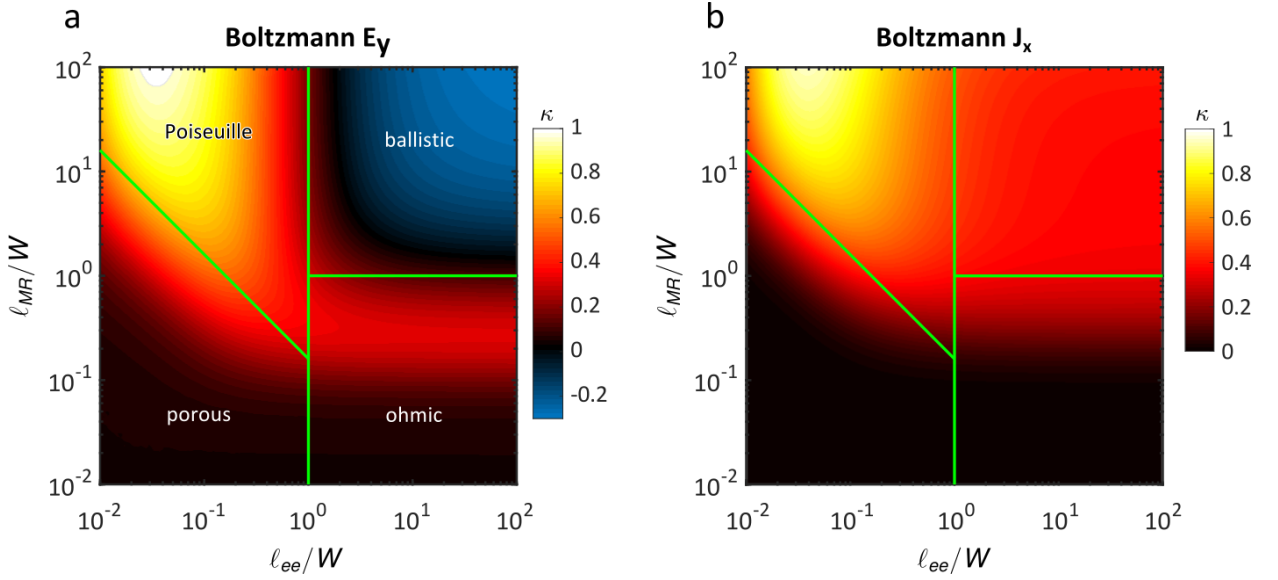

**Figure S7 Comparison of $E_y$ and $J_x$ from Boltzmann simulation. a**. Curvature $\kappa$ of $E_y$, as in fig 4b., calculated by Boltzmann simulation (section S5), and plotted on a logarithmic scale as a function of $l_{ee}/W$ and $l_{MR}/W$ for $W/R_c = 1.3$. Curvature is calculated over the center of the channel. Green lines: division into flow regimes as in fig. 4b. **b**. Curvature $\kappa$ of $j_x$ plotted for the same simulation as panel a. For $j_x$, the curvature in the ballistic regime is essentially constant at $\kappa \simeq 0.31$ and so the curvature of $j_x$ is less discriminating between the hydrodynamic and ballistic regimes than the curvature of $E_y$, which goes negative. In the other regimes, $j_x$ is very similar to $E_y$, and the differences between them diminish as each of the length-scales becomes much smaller than $W$. In the hydrodynamic regime the curvature saturates on the maximal possible value for a strictly parabolic profile, and in the porous regime it follows the length-scale $D_\nu = \sqrt{l_{MR} l_{ee}}$ as expected.